\documentclass[aps,prx,superscriptaddress,twocolumn,showpacs]{revtex4-1}
\usepackage[table]{xcolor}
\usepackage{pgf}
\usepackage[utf8]{inputenc}
\usepackage{amsmath}
\usepackage{braket}
\usepackage{amssymb}
\usepackage{amsmath}
\usepackage{bbold}
\usepackage{enumitem} 
\usepackage{appendix}
\usepackage{graphicx}
\usepackage{overpic}
\usepackage{epstopdf}
\usepackage[pdftex,bookmarks,colorlinks]{hyperref}
\usepackage{hyperref}

\usepackage{ulem}
\newcommand{\mean}[1]{\left\langle #1\right\rangle}
\newcommand{\nep}{\textrm{e}}
\newcommand\mdm[1]{\textcolor{black}{#1}}
\definecolor{caribbeangreen}{rgb}{0.0, 0.8, 0.6}
\newcommand\norm[1]{\left\lVert#1\right\rVert}

\begin{document}
%
%\title{Gauge time crystal}
\title{Homogeneous Floquet time crystal protected by gauge invariance}
\author{Angelo Russomanno}
\affiliation{Abdus Salam ICTP, Strada Costiera 11, I-34151 Trieste, Italy}
\affiliation{NEST, Scuola Normale Superiore \& Istituto Nanoscienze-CNR, I-56126 Pisa, Italy}
\affiliation{Max-Planck-Institut f\"ur Physik Komplexer Systeme, N\"othnitzer Strasse 38, D-01187, Dresden, Germany}

\author{Simone Notarnicola}
\affiliation{Dipartimento di Fisica e Astronomia ``Galileo Galilei'', via Marzolo 8, I-35131, Padova, Italy}
\affiliation{INFN, Sezione di Padova, via Marzolo 8, I-35131 Padova, Italy}

\author{Federica Maria Surace}
\affiliation{Abdus Salam ICTP, Strada Costiera 11, I-34151 Trieste, Italy}
\affiliation{SISSA, Via Bonomea 265, I-34136 Trieste, Italy}

\author{Rosario Fazio}
\affiliation{Abdus Salam ICTP, Strada Costiera 11, I-34151 Trieste, Italy}
\affiliation{Dipartimento di Fisica, Universit\`a di Napoli "Federico II", Monte S. Angelo, I-80126 Napoli, Italy}

\author{Marcello Dalmonte}
\affiliation{Abdus Salam ICTP, Strada Costiera 11, I-34151 Trieste, Italy}
\affiliation{SISSA, Via Bonomea 265, I-34136 Trieste, Italy}

\author{Markus Heyl}
\affiliation{Max-Planck-Institut f\"ur Physik Komplexer Systeme, N\"othnitzer Strasse 38, D-01187, Dresden, Germany}

\begin{abstract}
We show that homogeneous lattice gauge theories can realize nonequilibrium quantum phases with long-range spatiotemporal order protected by gauge invariance instead of disorder.
We study a kicked $\mathbb{Z}_2$-Higgs gauge theory and find that it breaks the discrete temporal symmetry by a period doubling.
In a limit solvable by Jordan-Wigner analysis we extensively study the time-crystal properties for large systems and further find that the spatiotemporal order is robust under the addition of a solvability-breaking perturbation preserving the $\mathbb{Z}_2$ gauge symmetry.
The protecting mechanism for the nonequilibrium order relies on the Hilbert space structure of lattice gauge theories, so that our results can be directly extended to other models with discrete gauge symmetries.
\end{abstract}
\maketitle
\textit{Introduction.--- }
Isolated quantum matter can feature phases with long-range order in highly excited states that cannot be captured by thermodynamic ensembles~\cite{2013Huse,2015Nandkishore}.
This crucially relies on ergodicity breaking and a failure of the Eigenstate Thermalization Hypothesis (ETH)~\cite{2016AdPhyETH}.
One robust mechanism for achieving such nonergodic behavior is to impose strong disorder giving rise to the many-body localized (MBL) phase~\cite{2015Nandkishore,2015Altman,Bloch1,2017Smith,2018Brenes,2018Smith}, which can host long-range ordered phases such as the MBL-spin glass~\cite{2013Huse,Pollmann14} or Floquet time crystals~\cite{vedika16,nayak,2017Zhang,2018Pal,Barrett_prl18,Barrett_PRB18}.
Recently, it has been realized that lattice gauge theories (LGTs) entail another robust mechanism for nonergodic dynamics in short-ranged systems protected by gauge invariance instead of disorder~\cite{2017Smith,2018Brenes,2018Smith} due to the specific structure of their Hilbert spaces, which are built up of disconnected superselection sectors~\cite{2018Brenes}. 
However, it has remained an open question to which extent they can also accommodate nonequilibrium phases with long-range order and therefore to which extent they can contribute to the open quest of realizing robust nonequilibrium ordered phases of homogeneous quantum \textcolor{black}{many-body systems}.%~\cite{2017Else,2019Yu,2019Schaefer,2019Choi,2019Iadecola,2018Huang,2018WangPoletti,2018GongUeda,2019Gambetta,2018OSullivan,2018Tucker,2019Zhu,2019Lazarides,6}.

\begin{figure}
	\begin{center}
           \begin{tabular}{c}
		\hspace{0.2cm}\fbox{\includegraphics[width=7.2cm]{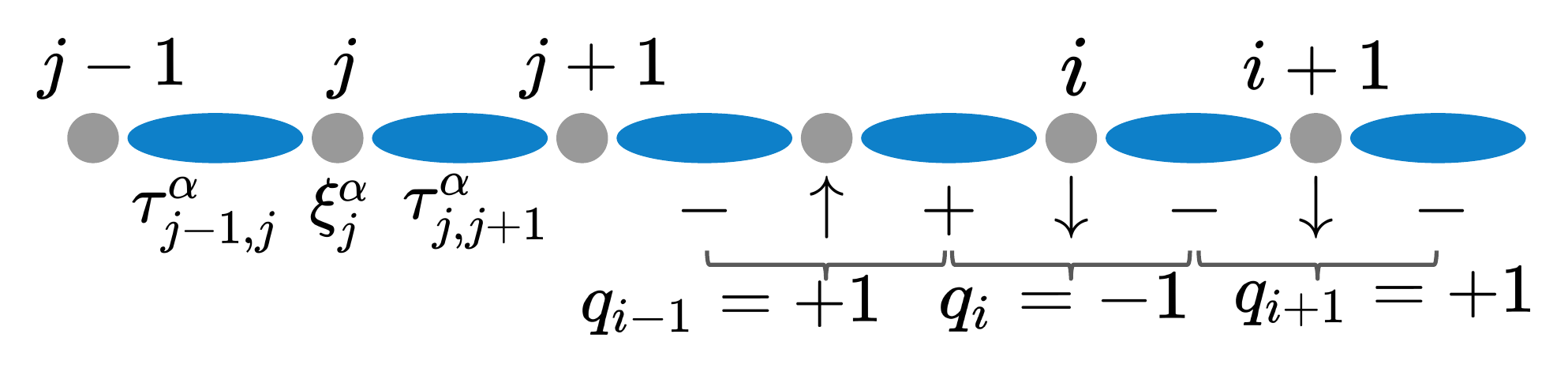}}\vspace{0.3cm}\\
		\includegraphics[width=8cm]{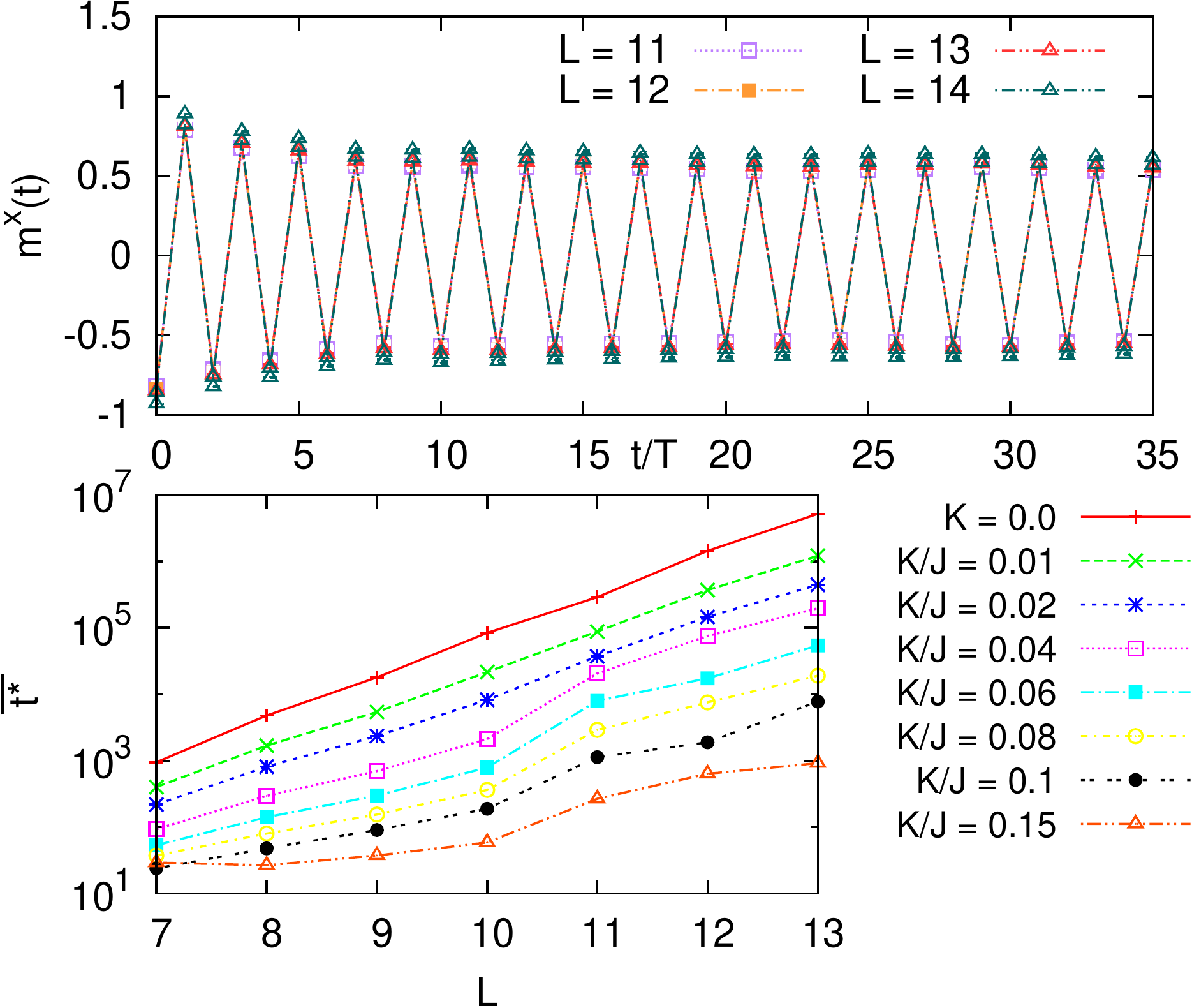}
           \end{tabular}
           		      \pgfputat{\pgfxy(-8.5,4.5)}{\pgfbox[left,top]{\footnotesize (a)}}
    \pgfputat{\pgfxy(-8.5,2)}{\pgfbox[left,top]{\footnotesize (b)}}
      \pgfputat{\pgfxy(-8.5,-1)}{\pgfbox[left,top]{\footnotesize (c)}}
		\caption{(a) Schematic illustration of the $\mathbb{Z}_2$ Higgs-LGT, with matter fields on the lattice sites $l$, represented by Pauli operators $\hat{\xi}_j^\alpha$, and gauge degrees of freedom by $\hat{\tau}^\alpha_{j,j+1}$ on the links. The local gauge symmetry imposes a locally conserved quantity given by the eigenvalues $q_i$ of the operator $\hat G_i=-\hat{\tau}^x_{i-1,i}\hat{\xi}^z_i\hat{\tau}^x_{i,i+1}$, which are included for a simple example, where $\uparrow \downarrow$ represents the eigenvalues of $\xi_j^z$ and $\pm$ of $\tau_{j,j+1}^x$, respectively.
			(b) Stroboscopic dynamics of the magnetization $m^x(t)$ of the gauge degrees of freedom in the kicked $\mathbb{Z}_2$ LGTs displaying period-doubling oscillations.
			(c) The decay time $\overline{t^\ast}$ of the period-doubling oscillations increases exponentially with system size $L$ marking the presence of a time-crystal behavior. We have taken a $\mathbb{Z}_2$-symmetry breaking initial state with $f=0.8$ [see discussion after Eq.~\eqref{asy:eqn}]. Numerical parameters: $\phi=1.02\pi$, $h/J=0.2$, $m/J=0.5$, $JT=1.0$, $N_{\rm real}=48$, and $K/J=0.1$ in (b). }
		\label{fig1}
	\end{center}
\end{figure}

\textcolor{black}{In this work we introduce a phase of quantum matter unique to LGTs that exhibits both spatial and temporal order thereby constituting a genuine nonequilibrium phenomenon. In particular, we show that homogeneous LGTs can feature robust time-crystalline phases in short-range systems protected by gauge invariance as opposed to previously studied cases that were relying on the presence of strong disorder. In order to realize such a 'gauge time crystal', we introduce a periodically kicked $\mathbb{Z}_2$ LGT which, as we find, displays a sub-harmonic response to the external drive associated with a period doubling, see Fig.~\ref{fig1}. We identify two necessary properties essential to realize a Floquet time crystal within the considered scheme: i) in a given superselection sector the LGT has to realize bond instead of field disorder in contrast to previously studied models of disorder-free localization~\cite{2017Smith,2018Brenes,2018Smith}; ii) the gauge symmetry has to be discrete and different from many previously studied nonergodic $U(1)$ LGTs~\cite{2017Smith,2018Brenes}.}
%~\cite{shapere_prl12,2019Zhu,Li_prl12,Wilczek_prl13,Bruno1,Ri_Bruno,Volovik2013,Bruno2,Nozi_res_2013,SachaTC,vedika16,vedika_K,Yao_prl,vedike_KK,biao,6,federica,2017Bernien,2017Zhang,Pal_prl18,Barrett_prl18,Barrett_PRB18,2018Tucker,fernando_prl}
%
%This ’gauge time crystal’ is characterized by both spatial and temporal long-range order.
%
%In this work we focus on the time-crystal physics~\cite{shapere_prl12,Li_prl12,Wilczek_prl13,Bruno1,Ri_Bruno,Volovik2013,Bruno2,Nozi_res_2013,SachaTC,vedika16,vedika_K,Yao_prl,vedike_KK,biao,6,federica,2017Bernien,2017Zhang,Pal_prl18,Barrett_prl18,Barrett_PRB18} and show that homogeneous LGTs can feature robust discrete time-crystalline phases protected by gauge invariance instead of disorder.
%
%We identify a period-doubling discrete Floquet time crystal in a kicked $\mathbb{Z}_2$ LGT (Fig.~\ref{fig1}) 
%
%This 'gauge time crystal' is 
%characterized by both spatial and temporal long-range order.
%
We solve the considered kicked $\mathbb{Z}_2$ LGT exactly by a mapping onto a free fermionic theory using a Jordan-Wigner (J-W) transformation, which allows us to explore the phase diagram for large system sizes.
We observe that the Floquet states appear in pairs with a quasienergy difference of $\pi$, so that our system shares many of the features of the $\pi$-spin glass in a periodically kicked Ising chain with quenched disorder~\cite{vedika16}.
Importantly, we find that this gauge time crystal represents a robust phase which does not require fine tuning and persists over a wide range of parameters.
In particular, we also study the influence of perturbations breaking the exact solvability and preserving the $\mathbb{Z}_2$ gauge symmetry, where we find numerical evidence for stability by means of exact diagonalization. \textcolor{black}{%We stress again the importance of the discrete symmetry: i
We discuss how to extend our analysis to a $\mathbb{Z}_N$-symmetric LGT along the lines of~\cite{federica}.} %on the {\it bonds}.
%{This is very important to break the degeneracies in the quasienergy spectrum which would make the spatio-temporal long-range order unstable to perturbations; the on-site effective disorder of the known ergodicity-breaking LGT would not be enough for that~\cite{2018Brenes}.}
%
The mechanism behind this time-crystalline phase relies on gauge invariance and can therefore be directly extended to other LGTs with discrete gauge symmetries. Importantly, our observation of a robust time-crystalline phase in a homogeneous short-ranged system goes beyond recent approaches which lead to prethermal spatiotemporal order~\cite{2017Else,2019Yu,2019Schaefer,2019Choi,2019Iadecola}, and dissipative dynamics~\cite{2019Gambetta,2018WangPoletti,2018Tucker,2018GongUeda,2018OSullivan,2019Zhu,2019Lazarides}.

\textit{The model.--- }
We consider a $\mathbb{Z}_2$ Higgs-LGT in one spatial dimension. The theory describes the dynamics of Higgs fields - defined by Pauli-matrix operators $\hat{\xi}^\alpha_j$ at vertex $j$ on the lattice - coupled to $\mathbb{Z}_2$ gauge fields - defined by $\mathbb{Z}_2$ parallel transporters $\hat{\tau}^x_{j,j+1}$ at the bond $(j, j+1)$ as illustrated in Fig.~\ref{fig1}(a). The system Hamiltonian reads~\cite{creutzbook,fradkinbook}
 \begin{align}\label{HLGT}
\hat H_0=&\, \frac{m}{2}\,\sum_{j=1}^L\hat{\xi}_j^z + J\sum_{j=2}^{L-1}\hat{\tau}_{j-1,\,j}^x\hat{\tau}_{j,\,j+1}^x +h\sum_{j=1}^{L-1}\hat{\xi}_j^x\hat{\tau}_{j,\,j+1}^z\hat{\xi}_{j+1}^x \, .
\end{align}
%
%Here, the Pauli matrices $\xi_j$ represent the Higgs fields, which 
The Higgs-field operators can also be interpreted as hard-core bosons $\hat{b}_j$ with $\hat{\xi}^x_j = \hat{b}^\dagger_j+\hat{b}_j$. The first two terms denote mass and gauge interactions, while the third describes the coupling between the Higgs and gauge fields. We drive the $\mathbb{Z}_2$ Higgs-LGT out of equilibrium by periodically kicking the strength of the Higgs-gauge coupling, leading to the following time-dependent Hamiltonian
\begin{equation} \label{kickedHLGT}
\hat H(t) = \hat H_0 + \frac{\phi}{2} \sum_{n=-\infty}^{+\infty}\delta(t-nT)\sum_{j=1}^{L-1}\hat{\xi}_j^x\hat{\tau}_{j,\,j+1}^z\hat{\xi}_{j+1}^x
\end{equation}	
This system exhibits a local symmetry: $\hat H(t)$ commutes with the operators $\hat G_j = -\hat{\tau}_{j-1,\,j}^x\hat{\tau}_{j,\,j+1}^x \hat{\xi}_j^z$ (which can be understood as the complex exponentials of the local Gauss' operators). Thus, the Hilbert space of size $2^{2L-1}$ is partitioned in $\mathcal{N}=2^L$ superselection sectors, where all the states $|\Psi_{\{\ q_\alpha\} }\rangle$ in a given sector are identified by the same set of local static charges $q_j=\pm1$ via $\hat G_j |\Psi_{\{\ q_\alpha\} }\rangle = q_j|\Psi_{\{\ q_\alpha\} }\rangle$. 
In the following we consider initial product states of the form {$|\Psi\rangle = |\varphi \rangle_H \otimes | \psi \rangle_g$}% $|\Psi\rangle = |+\rangle_H \otimes | \psi \rangle_g$
where {$|\varphi\rangle_H$} %$|+\rangle_H$
%denotes a fully polarized state for the Higgs fields with $\hat\xi_j^x |+\rangle_H = |+\rangle_H$
{is a product state which satisfies ${}_H\langle\varphi| \hat \xi_j^z|\varphi\rangle_H=0$}
for all $j=1,\dots,N$ and $|\psi \rangle_g$ is the initial condition for the gauge degrees of freedom, that we will specify later in the text. 
%We note that this choice of initial state is rather generic, as in Ref.~\cite{2018Brenes}; below, we choose specific, simple forms of $|\psi\rangle_g$ for the sake of a clearer interpretation of the dynamics.%\textcolor{black}{(Maybe this is true also for all the possible superpositions of states of the form $|\Psi\rangle = |\{\sigma_k\}\rangle_H \otimes | \psi \rangle_g$ where $\hat\xi_j^x |\{\sigma_k\}\rangle_H = \sigma_j|\{\sigma_k\}\rangle_H$ and $\sigma_k=\pm 1$ $\forall k$. Am I wrong?)}
%
Such initial conditions, which represent superpositions over many superselection sectors, can yield robust nonergodic behavior for LGTs and disorder-free localization~\cite{2017Smith,2017Smith2,2018Brenes,2018Smith}.
Concretely, for our $\mathbb{Z}_2$ LGT the dynamics in a given superselection sector specified by the charges $\{ q_\alpha \}$ is determined by an effective Hamiltonian %(see Appendix~\ref{derivation:sec})
%
%It has recently been shown that the dynamics in LGTs for initial states $|\Psi\rangle=\sum_{\{ q_\alpha \}} C_{\{q_\alpha\}} |\Psi_{\{\ q_\alpha\} }\rangle$,  which are superpositions over different superselection sectors, can yield robust nonergodic dynamics.
%
%For generic observables $\mathcal{O}$, which don't couple the different sectors $\{q_\alpha\}$, the time evolution yields $\langle \mathcal{O}(t) \rangle = \sum_{ \{ q_\alpha \} } |C_{ \{ q_\alpha \} }|^2 \langle \Psi_{\{\ q_\alpha\} }(t)| \mathcal{O} | \Psi_{\{\ q_\alpha\} }(t) \rangle$ resembling an effective disorder average  over different inhomogeneous superselection sectors that can lead to disorder-free localization.
%
%Here, we consider initial product states of the form $|\Psi\rangle = |+\rangle_H \otimes | \psi \rangle_g$ where $|+\rangle_H$ denotes a fully polarized state for the matter fields with $\xi_j^x |+\rangle_H = |+\rangle_H$ for all $j=1,\dots,N$ and $|\psi \rangle_g$ the initial condition for the gauge degrees of freedom, that we will specify later in the text.
%
%Using $\xi_j^z = -q_j \tau_{j-1,\,j}\tau_{j,\,j+1}$ it is possible to integrate out the Higgs field exactly for the Hamiltonian in Eq.~(\ref{kickedHLGT}) in a given superselection sector. Redefining the gauge fields according to $\tau_{j,j+1} \mapsto $ one finally obtains:
 \begin{align} \label{Hamour:eqn}
\hat H_{\{q_\alpha\}}(t)& =\, \sum_{j=2}^{L-1}J_j\hat{\tau}_{j-1,\,j}^x\hat{\tau}_{j,\,j+1}^x+h(t)\sum_{j=1}^{L-1}\hat{\tau}_{j,\,j+1}^z\,,
\end{align}
with $h(t) = h + (\phi/2) \sum_{n=-\infty}^{+\infty}\delta(t-nT)$, $J_j=[1-q_j m/(2 J)]$ and \textcolor{black}{the $\hat{\tau}_{j,\,j+1}^\alpha$ operators redefined with respect to Eqs.~\eqref{HLGT} and~\eqref{kickedHLGT} (see Appendix~\ref{derivation:sec} for details).}
This integration is related to the duality between Ising models and Ising LGTs~\cite{Wegner1971,McCoy1983}.
As a result the Hamiltonian becomes a kicked transverse-field Ising chain with binary bond disorder due to $q_j = \pm 1$, which can be solved exactly via a J-W transformation for large systems. \textcolor{black}{We emphasize that}\textcolor{black}{, due to the presence of degeneracies in the unperturbed Floquet spectrum, it is a priori less clear whether bond disorder - with respect to one with a continuous distribution - is able to induce MBL in order to get a time crystal.}
%{ it is impossible to predict a priori if this kind of disorder is able to induce MBL and if it can break the degeneracies in the Floquet spectrum in order to get a robust time crystal (we will find a positive answer to both questions).}
%
We will also study the influence of a perturbation of the form $ \hat H_K = 4K\sum_{j=2}^{L-1}{\hat{\xi}_{j-1}^x}\hat{\tau}_{j-1,j}^z\hat{\tau}_{j,j+1}^z{\hat{\xi}_{j+1}^x}$ breaking the J-W solvability. After the integration it adds a transverse interaction term for the gauge fields
\begin{equation}\label{hak:eqn}
	\hat{H}_{\{ q_\alpha \}}^K(t) = \hat H_{\{q_\alpha\}}(t) + 4K\sum_{j=2}^{L-1} \hat{\tau}_{j-1,\,j}^z \hat{\tau}_{j,j+1}^z \, .
\end{equation}
We solve the dynamics of the LGT in a set of $N_{\rm real}$ randomly chosen superselection sectors and finally perform an average when computing observables.
In the shown data we include error bars resulting from the finiteness of $N_{\rm real}$.
But let us emphasize again that the overall problem is homogeneous both in the initial condition and in the Hamiltonian.

\textit{Initial conditions and observables.--- }
In order to reveal both the temporal and spatial order we use two complementary setups.
On the one hand, we take initial conditions which explicitly break the $\mathbb{Z}_2$ symmetry of the model yielding a nonzero magnetization $m^x$ for the gauge degrees of freedom which we then monitor in the subsequent evolution:
\begin{equation}\label{maga:eqn}
  m^x(t)=\frac{1}{L-1}\sum_{j=1}^{L-1}\langle \hat{\tau}_{j,\,j+1}^x \rangle_t\,,
\end{equation}
where we have defined $\mean{\cdots}_t\equiv\overline{_g\bra{\psi(t)}\cdots\ket{\psi(t)}_g}$ and the overline marks the average over the $N_{\rm real}$ pseudo-disorder realizations~\cite{media_not}. In this way we obtain direct access to the time-crystalline period-doubling dynamics. % anticipating some of the results discussed in following sections.
%{%We would like to show to the reader the time-crystal period-doubling dynamics: Anticipating a little bit,
In Fig.~\ref{fig1}(b) we show results for $m^x(t)$ in the fully interacting case $K\neq 0$ obtained through exact diagonalization. We see the existence of period-doubling oscillations which are {\it persistent} for an infinite time in the thermodynamic limit. We show this fundamental property of persistence~\cite{nayak} in Fig.~\ref{fig1}(c), where we see that the decay time $\overline{t^*}$ of the period doubling oscillations exponentially scales to infinity with the system size. We determine $\overline{t^*}$ as the time after which $(-1)^{t/T}m^x(t)$ changes sign~\cite{nayak1, federica} averaged over disorder. %Let us now move to discuss the time-crystal behaviour and its robustness in different regimes.}%we can access the temporal long-range order, as we show in Fig.~\ref{fig1}c.

On the other hand we can choose initial conditions which are $\mathbb{Z}_2$-symmetric with a vanishing magnetization $m^x(t)$, which allows us to address the spatial long-range ordering in the system. For that purpose we study the correlation parameter
\begin{equation} \label{corrpa:eqn}
\mathcal{S}_t^{xx}=\frac{1}{(L-1)(L-2)}\sum_{i,j=1,(i\neq j)}^{L-1}\langle \hat{\tau}_{j,\,j+1}^x\hat{\tau}_{i,\,i+1}^x\rangle_t\,,
\end{equation} 
with $\mean{\cdots}_t$ defined as above. Whenever $\mathcal{S}_t^{xx}>0$ while at the same time $m^x(t)=0$, the system exhibits long-range spatial order.
%{We now want to illustrate the time-crystal period-doubling dynamics, anticipating some of the results discussed in following sections.}
%{%We would like to show to the reader the time-crystal period-doubling dynamics: Anticipating a little bit,
%Let us consider an example of evolution of $m^x(t)$ in the fully interacting case, obtained through exact diagonalization (upper panel of Fig.~\ref{fig1}). We see the existence of period-doubling oscillations which are {\it persistent} for an infinite time in the thermodynamic limit. We show this fundamental property of persistence~\cite{nayak,biao} in the lower panel of Fig.~\ref{fig1}, where we see that the decay time $\mean{t_d}$ of the period doubling oscillations exponentially scales to infinity with the system size. (We evaluate $\mean{t_d}$ as the average over disorder of the {time, in units of $T$, after which} $(-1)^{t/T}m^x(t)$ changes sign.) Let us now move to discuss the time-crystal behaviour and its robustness in different regimes.}

\textit{{Exactly solvable case.--- }}
Let us first focus on the case with $K=0$, where the model can be mapped onto a system of non-interacting fermions by means of a J-W transformation. In each superselection sector $\{q_\alpha\}$ we initialize the dynamics with the same initial state $\ket{\psi}_g$ chosen as the ground state of the Hamiltonian $\hat H_0=\sum_{j=2}^{L-1}\hat\tau_{j-1,\,j}^x\hat\tau_{j,\,j+1}^x+h_0\sum_{j=1}^{L-1}\hat\tau_{j,\,j+1}^z$. This state has a non-vanishing correlation parameter if $h_0<1$ and is symmetric under $\mathbb{Z}_2$ which allows us to address the long-range spatial ordering in the system; for a study of the temporal order we perform a spectral analysis, as we are going to detail below. In the J-W framework it is well known how to numerically study the dynamics and how to evaluate the correlation parameter as a Pfaffian %and we refer the reader to the appropriate literature
(see~\cite{lieber,pfeuty,russ1,santoro,barouch}). Here it is enough to say that the dynamics is induced by an effective $2(L-1)\times 2(L-1)$ time-periodic single-particle Hamiltonian. This is important to mention because we can compute the $2(L-1)$ single-particle Floquet states and the $2(L-1)$ single-particle quasi-energies $\epsilon_\alpha$ (see for instance~\cite{emanuele}). These quantities will play an important role in what follows.
\begin{figure}
  \begin{center}
    \includegraphics[width=8cm]{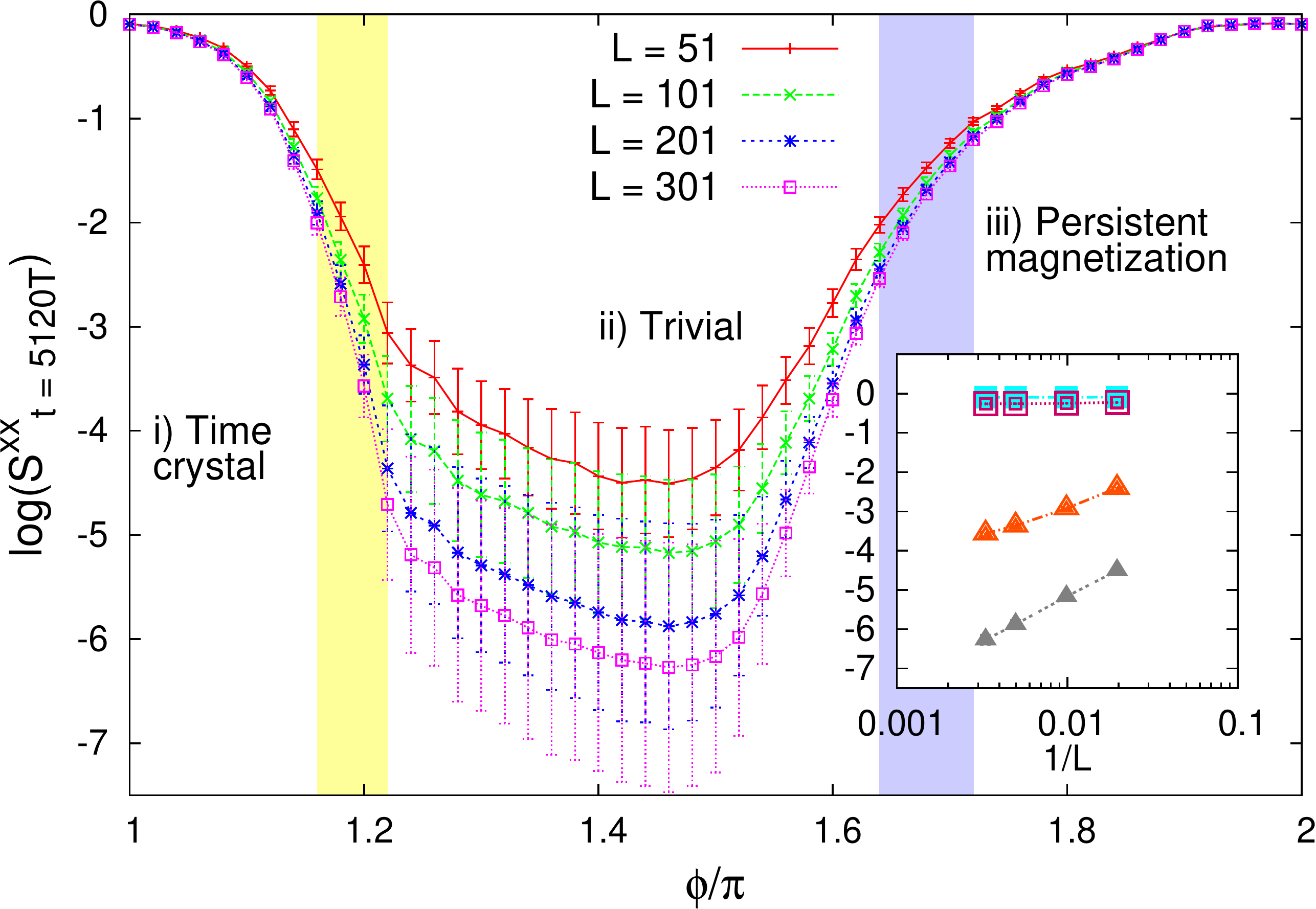}
\caption{Spatial long-range order in the exactly solvable limit $K=0$: Asymptotic long-time value $\mathcal{S}_{\mathrm{asy}}^{xx}$ versus $\phi$ for different values of system size $L$. (Inset) System-size dependence of $\mathcal{S}_{\mathrm{asy}}^{xx}$ from top to bottom for $\phi/\pi=1,\,1.06,\,1.2,\,1.4$. Numerical parameters as in Fig.~\ref{fig1}, except $K=0$ and $N_{\rm real}\ge104$.}
\label{fig2_comp}
  \end{center}
\end{figure}

\begin{figure}
  \begin{center}
    \includegraphics[width=8cm]{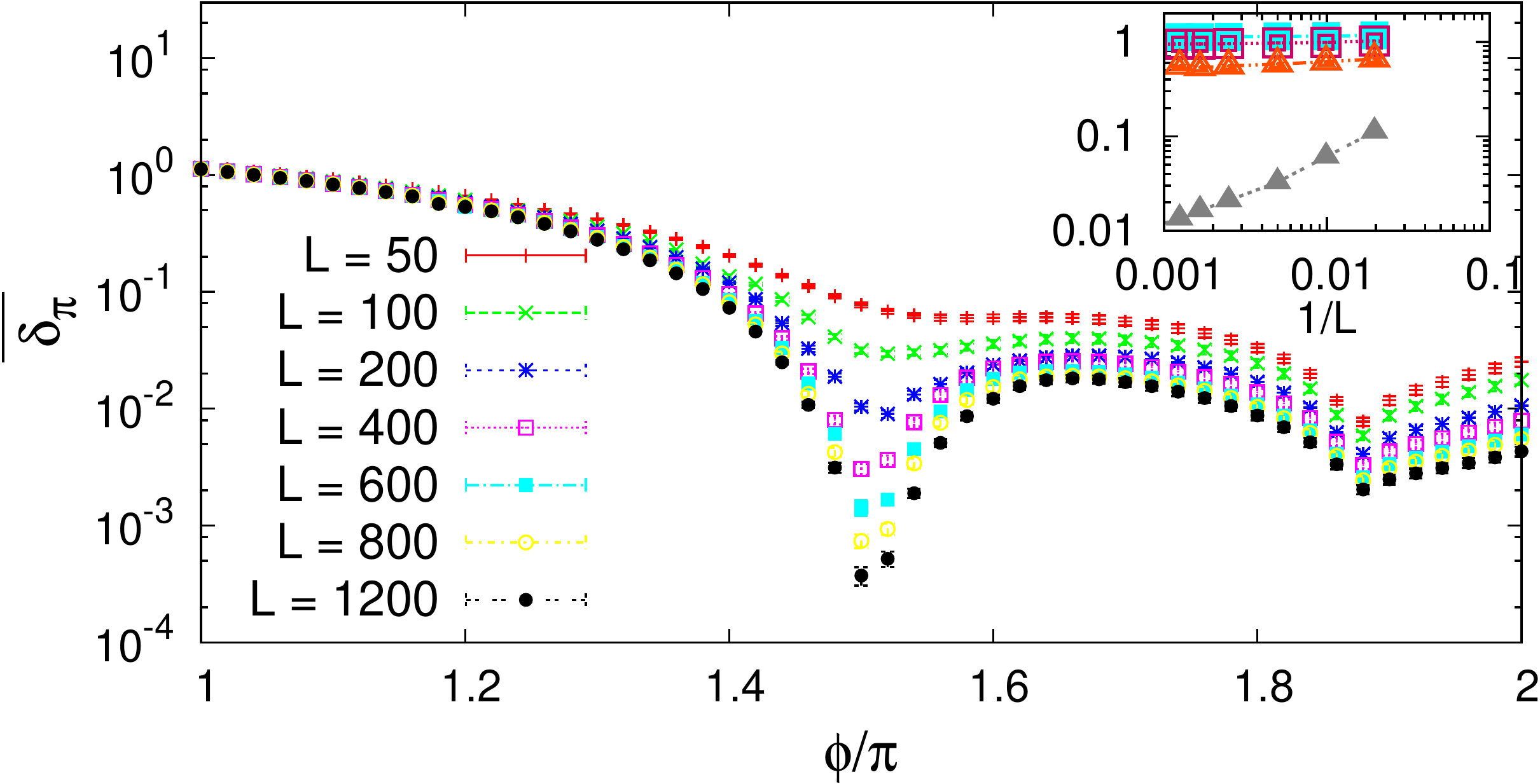}
\caption{Disorder-averaged single-particle Floquet spectral gap $\overline{\delta_\pi}$ as a function of $\phi$ for different $L$ using the same parameters as in Fig.~\ref{fig2_comp}. (Inset) System-size dependence of $\overline{\delta_\pi}$.
}
\label{fig2b}
  \end{center}
\end{figure}
% (Lower panel) 

%As we can see in the inset of Fig.~\ref{fig2_comp}, 
We find that the correlation order parameter reaches an asymptotic value $\mathcal{S}_{\rm asy}^{xx}$ after a transient (see the discussion below Eq.~\eqref{asy:eqn}). We plot the long-time value of $\mathcal{S}_t^{xx}$ as a function of kicking strength $\phi$ for different values of $L$ in the main panel of Fig.~\ref{fig2_comp}. We observe three regimes whose separating phase boundaries we indicate by the colored zones. In the regimes {\it i)} and {\it iii)} $\mathcal{S}_{\rm asy}^{xx}$ converges to a nonzero value as $L\to\infty$, while in regime {\it ii)} $\mathcal{S}_{\rm asy}^{xx}$ vanishes as the $L$ is increased (see also the inset of Fig.~\ref{fig2_comp}). Both regions {\it i)} and {\it iii)} mark the existence of an eigenstate phase~\cite{2013Huse,2015Nandkishore}, where eigenstates exhibit long range spatial order (as in~\cite{Pollmann14}, for instance). \textcolor{black}{This eigenstate order is protected by disorder and MBL since in a clean short-range one-dimensional spin interacting thermalizing system with $\mathbb{Z}_2$ symmetry such order is impossible (this result is easily shown for clean $\mathbb{Z}_2$ one-dimensional spin chains~\cite{scalettar}, where long-range order is possible only in the ground state)}. % due to the Mermin-Wagner theorem.

Although the behaviour of $\mathcal{S}_{\rm asy}^{xx}$ is qualitatively similar in both {\it i)} and {\it iii)}, these two regions mark different phases since {\it i)} in addition also supports temporal order. An example of this property for $\phi=1.02\pi$ can be seen in Fig.~\ref{fig1}(c) (curve with $K=0$): The system is initialized in a state explicitly breaking the $\mathbb{Z}_2$ symmetry and the decay time $\overline{t^*}$ exponentially increases with the system size. % as we will now discuss. This property can be determined in two ways. First, by initializing the system in state the breaks the $\mathbb{Z}_2$ symmetry, the dynamics of the magnetization $m^x(t)$ [Eq.~\eqref{maga:eqn}] we observe that in {\it i)} $m_x(t)$ exhibits a period doubling dynamics as shown in Fig.~\ref{fig1} absent in {\it iii)}. Second, t
This fact can be understood by an analysis of the Floquet spectrum~\cite{vedika16}.
%... We claim that in the regime {\it i)} it changes sign at each kick and then oscillates with period $2T$, giving rise to a period-doubling time crystal, while in the regime {\it iii)} it is constant with time. %In both these conditions the correlation order parameter reaches an asymptotic value and we cannot use it to tell what is the situation. 
%In order to support this claim we must consider the single-particle Floquet spectrum which we have introduced above. As shown in~\cite{vedika16}, in order to have a period-doubling time crystal
The presence of a temporal time-crystal ordering corresponds to spectral pairing, where each Floquet state has a partner with quasienergy shifted by $\pi$. This situation is realized if there is a single-particle quasienergy exactly at $\pi$ with a marked gap separating it from the rest of the spectrum. In this way it does not hybridize with the bulk, and each many-body Floquet state has a $\pi$-shifted partner obtained by adding the quasiparticle with quasienergy $\pi$. We evaluate this gap as $\overline{\delta_\pi}=\frac{1}{N_{\rm real}}\sum_{q=1}^{N_{\rm real}}\left[\epsilon_{2L-2}^{(q)}-\epsilon_{2L-3}^{(q)}\right]$~\cite{notquasi} and plot it in Fig.~\ref{fig2b}. We see that it is non-vanishing in all the regime {\it i)}. Moreover, as we show in Appendix~\ref{single:sec}, in this regime $\epsilon_{2L-2}$ averaged over the disorder is exactly equal to $\pi$. In Appendix~\ref{single:sec} we show also that the single-particle bulk Floquet states are always Anderson localized. This is very important, because without localization {it is possible to have a gap in the Floquet spectrum at $\pi$ and still observe no time crystal (see for instance~\cite{emanuele}): In the absence of localization,}
local operators expand in time obeying the Lieb-Robinson bound and no time-periodic behaviour whatsoever is possible~\cite{vedika_K}. %For instance, in~\cite{emanuele} the authors find a gap in the Floquet spectrum at $\pi$ but no time-crystal, because of the absence of localization.
{Of course, the transition to localization and the one to glassy order of the excited eigenstates are independent~\cite{Pollmann14}, {and this is the reason} why the transition from regime {\it i)} and {\it ii)} occurs at a value of $\phi$ different from the one where $\overline{\delta_\pi}$ vanishes.} In Fig.~\ref{fig2_comp} we have initialized with a specific value of $h_0$, but we have checked that the presented phenomenology doesn't depend on this choice.

\begin{figure}
  \begin{center}
    \includegraphics[width=8cm]{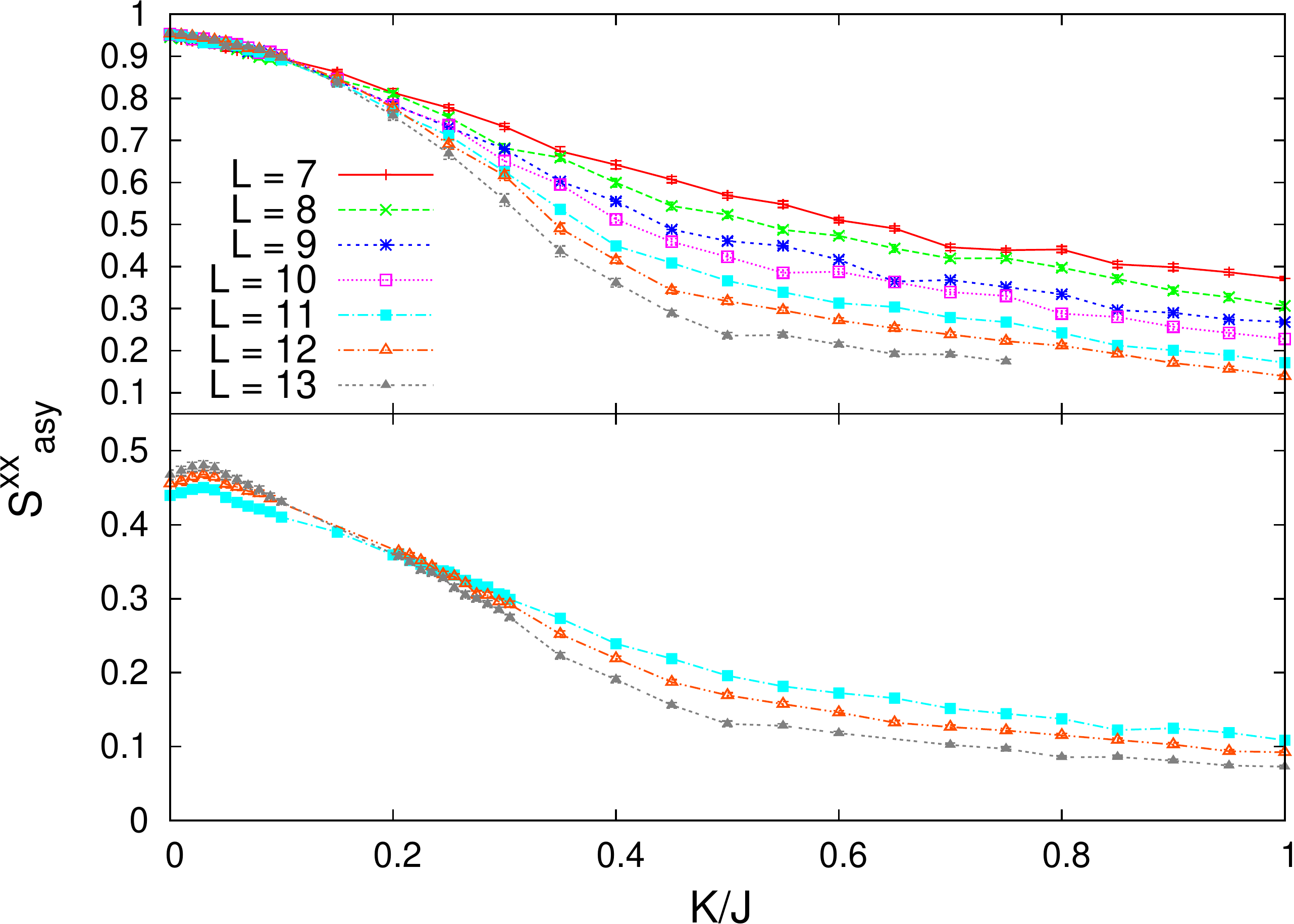}
\caption{Stability of the time crystal: $\mathcal{S}_{\rm asy}^{xx}$ as a function of $K$ for various $L$ and two different initial states with $f=1$ (upper panel) and $f=0.8$ (lower panel). Numerical parameters as in Fig.~\ref{fig1}.}
\label{fig3}
  \end{center}
\end{figure}

\textit{{General case.--- }}
%Now we have demonstrated the existence of a time-crystal regime in the integrable limit which is robust and exists for an interval of parameters.  W
At this point we break J-W solvability by considering the term of Eq.~\eqref{hak:eqn}, with $K\neq0$. %put an integrability-breaking term in the Hamiltonian setting $K\neq 0$ in the Hamiltonian of Eq.~\eqref{Hamour:eqn}. %We remark that this term does not break the $\mathbb{Z}_2$ symmetry of the Hamiltonian.  we can no more resort to the J-W transformation but
%In order to see the persistence of the time crystal, we start considering 
We consider a value of $\phi$ for which we see this phenomenon at $K=0$; then we take $K\neq 0$ and we study the properties of the asymptotic correlation parameter. %We see if it scales to 0 as we increase $L$, or there is a
An interval of $K$ where this quantity does not scale with the size would mark the persistence of the time crystal. %Now {we need to do a normal exact diagonalization in order to perform a numerical analysis}: Due to the limits of our computational resources, we cannot consider sizes larger than $\sim 13$ sites. 
We now perform a conventional exact-diagonalization simulation of the system, up to size $L=13$. To evaluate the asymptotic correlation parameter, we can resort to the FLoquet diagonal ensemble and we get
\begin{equation}\label{asy:eqn}
  \mathcal{S}_{\rm asy}^{xx} = \sum_{i,j=1,(i\neq j)}^{L-1}\sum_{\beta=1}^{\mathcal{N}}\frac{\overline{|R_\beta|^2{_g\!\bra{\phi_\beta}}\hat\tau_{j,\,j+1}^x\hat\tau_{i,\,i+1}^x\ket{\phi_\beta}_g}}{(L-1)(L-2)}\,,
\end{equation}
where $\ket{\phi_\beta}_g$ are the many-body Floquet states, $\mathcal{N}$ is the dimension of the Hilbert space and $R_\beta\equiv{_g\left\langle\psi(0)\right.}\ket{\phi_\beta}_g$ denotes the overlap with the initial state.
%The destructive interference killing the off-diagonal non-degenerate terms in the long time is provided by the disorder average.
We remark that we can use Eq.~\eqref{asy:eqn} even if the many-body Floquet quasienergies $\mu_\beta$ appear in degenerate
pairs, due to the $\mathbb{Z}_2$ symmetry. The point is that the operators $\hat\tau_{j,\,j+1}^x\hat\tau_{i,\,i+1}^x$ %are degenerate in the two-dimensional degenerate Floquet subspaces, due to
{commute with} the same $\mathbb{Z}_2$ symmetry {and hence have no matrix elements between states with different parity} (the detailed demonstration along the lines of~\cite{io} is in Appendix~\ref{convergence:sec}). We plot the dependence of $\mathcal{S}_{\rm asy}^{xx}$ versus $K$ for different $L$ in Fig.~\ref{fig3}. We take two different initial conditions, in the upper panel we take the state with all the spins pointing down along the $x$ axis ($\ket{\psi}_g=\ket{s_{1,2}^x=-1\,\ldots\,s_{j,j+1}^x=-1\,\ldots\,s_{L-1,L}^x=-1}_g$), in the lower panel we take the uniform superposition of all the eigenstates of $\hat\tau_{j,\,j+1}^x\;\forall\,j$ obeying the condition $\sum_{j=1}^{L-1}s_{j,j+1}^x\leq-(L-1)f$ with $f=0.8$. We see that for $K\lesssim 0.2$ there is no decrease with $L$, marking the persistence of the time-crystal behaviour. This persistence can be seen also in Fig.~\ref{fig1}(c) where the $\overline{t^*}$ introduced above exponentially increases with $L$.

\paragraph*{Time crystallinity in Abelian lattice gauge theories.--- } %While the discussion so far focused exclusively on the $\mathbb{Z}_2$ case, w
We now investigate more generally if time crystallinity can appear in disorder-free Abelian LGTs in (1+1)-d. We consider the generic Hamiltonian coupling Higgs fields to {Abelian gauge fields}~\cite{kogut1975hamiltonian}:
\begin{eqnarray}
\hat{H} &=&  m\sum_{j=1}^{L} (-1)^j \hat{n}_j+ \sum_{j=1}^{L-1}(\hat{\varphi}^\dagger_j \hat{\mathcal{U}}_{j,j+1}\hat{\varphi}_{j+1} +\text{h.c.})\nonumber\\
&+& \frac{g^2}{2}\sum_{j=1}^{L-1}\hat E_{j,j+1}^2 + \hat{H}(t)\,. %\textcolor{black}{\text{How is $\hat{H}(t)$ defined?}}
\end{eqnarray}
where $\hat n_j=\hat\varphi^\dagger_j \hat\varphi_j$ is the Higgs occupation on site $j$ and $\hat E_{j,j+1}, \hat{\mathcal{U}}_{j,j+1}$ are respectively the electric field and the parallel transporter, and $\hat{H}(t)$ is defined analogously to the $\mathbb{Z}_2$ case above. The electric-field interaction energy is local in these theories, differently from the $\mathbb{Z}_2$ term involving at least two neighboring sites. For a $\mathbb{Z}_N$ LGT
% we consider the time evolution starting from
% \begin{equation}
%  |\Psi\rangle_N^0 = \frac{1}{\mathcal{N}}\sum_{\alpha_{12}=-N/2}^{N/2-1}.. \sum_{\alpha_{L-2, L-1}=-N/2}^{N/2-1}|0, \alpha_{12}, 1, \alpha_{23} ..\rangle
%  \end{equation}
% which corresponds to an initial state with the gauge fields in an equal weight superposition of all possible eigenvalues of the electric field, and a product state of the matter fields ($\mathcal{N}$ is a normalization factor)
(i.e. a theory where now $\hat{\mathcal{U}}_{j,j+1}$ and $\hat E_{j,j+1}$ are not Pauli matrices but the more general clock operators), we can use a similar approach as the one used in the $\mathbb{Z}_2$ LGT. We consider an initial state where matter is in an equal-weight superposition of all possible eigenvalues of the Higgs number operator, and the gauge fields are in a generic state.
The evolution of such states can be mapped exactly into the one of $\mathbb{Z}_N$ clock models under the effect of quasi-random local fields: since the latter class of models has been shown to display time-crystal behavior for small values of $N$ and random disorder~\cite{federica}, it is natural to expect that the mechanism discussed above holds true also for $N>2$. This mechanism does not work for continuous $U(1)$ LGTs (see Appendix~\ref{gauge:sec} for details), which, however, doesn't exclude other ones for the generation of time crystals in such theories.

%\mdm{The case of a continuous gauge symmetry shall instead be handled with care. As we discuss in the SM, the approach outlined above does not lead to time-crystalline behaviour for $U(1)$ LGTs.}

%

\textit{{Concluding discussion.---}}
In this work we have demonstrated that homogeneous LGTs can realize time-crystal phases, where the protecting nonergodicity is enforced by the local constraints imposed by gauge invariance.
%
%The absence of thermalization avoids consequently the applicability of no-go theorems for time crystalline behavior in equilibrium states of short-ranged systems.~\mh{[Here we could take some of the references that have been in the introduction before.]}
%
In more general terms, our results show that homogeneous LGTs can realize eigenstate order, which naturally leads to the question to which extent also other eigenstate phases can occur in homogeneous LGTs, e.g., analogues of the MBL-spin glass~\cite{2013Huse,Pollmann14} or topological order at elevated energy densities~\cite{bahri2015localization}.
\mdm{Our results are of immediate relevance to experiments realizing lattice gauge theory dynamics~\cite{Wiese:2013kk,Dalmonte:2016jk,Zohar2015} in both trapped ions~\cite{ExpPaper} and cold atom systems~\cite{Bernien2017,Surace:aa}. In particular, scalable proposals have been formulated~\cite{Zohar:2017aa,Notarnicola:aa}, and several experiments have already demonstrated the building blocks~\cite{Anderlini_2007,Trotzky:2008aa,Schweizer:aa,Mil:aa} for discrete lattice gauge theories of relevance to gauge time crystals.}

Further, our results can be directly extended to $\mathbb{Z}_N$ LGTs which opens up the possibility, in principle, of generating period {$N$}-tupling time-crystals.
While our approach cannot be immediately applied to LGTs with continuous groups, it would be intriguing to see whether discrete non-Abelian symmetries can also support the formation of defect-free time crystals.

\begin{acknowledgements}
{We acknowledge fruitful discussions with R.~Khasseh and M.~Wauters. A.~R. and R.~F. thank the European Union for partial financial support through QUIC project (under Grant Agreement 641122). A.~R. thanks the Max-Planck-Institut f\"ur Physik Komplexer Systeme for partial financial support and the warm hospitality received during the preparation of this work. {S.~N. and M.~D. acknowledge partial support by the H2020 Project - QUANTUM FLAGSHIP - PASQUANS (2019-2022)}. We are indebted with G.~E. Santoro for the subroutine performing the diagonalization of unitary operators.
This work is partly supported by the ERC under grant number 758329 (AGEnTh) and by the QUANTERA project QTFLAG.}
\end{acknowledgements}
%
%\bibliography{../bibliography,paper_06Notes_mdm}

\begin{thebibliography}{10}
\bibitem{media_not}
The system naturally averages over the pseudo-disorder realizations, so the
  errorbars we consider are the fluctuations of the average. They are evaluated
  as the standard deviation over disorder of the considered quantity divided by the
  square root of the number of realizations.
\bibitem{notquasi}
The quasienergies are in increasing order. If there is a spectral gap,
  $\epsilon_{2L-2}$ is the one at $\pi$. In this case, due to the particle-hole
  symmetry, $\epsilon_1=-\pi$.
\end{thebibliography}

%
\newpage
\appendix
\section{Derivation of the effective Hamiltonian $\hat H_{\{q_\alpha\}}(t)$}\label{derivation:sec}
The derivation of the effective Hamiltonian $\hat H_{\{q_\alpha\}}(t)$ from  $\hat H(t)$, defined respectively in Eqs.~(3)  and~(2) of the main paper, needs two steps.
In the first, we restrict ourselves to one of the superselection sectors defined by a set of static charges $\{q_\alpha\}$.
{To do so, for a generic state $|\Psi\rangle$, we consider its component $|\Psi_{\{q_\alpha\} }\rangle$ on the chosen sector, defined as
\begin{equation}
 |\Psi_{\{q_\alpha\} }\rangle= \frac{P_{\{q_\alpha\}}|\Psi\rangle}{\norm{P_{\{q_\alpha\}}|\Psi\rangle}}
\end{equation}
where $P_{\{q_\alpha\}}=\prod_j P_j(q_j)$ is the projector on the chosen superselection sector and $P_j(q_j)=(1-q_j\hat{\tau}_{j-1,\,j}^x\hat{\tau}_{j,\,j+1}^x \hat{\xi}_j^z)/2$ projects on the sector with static charge $q_j$ on site $j$.
}%: As stated before,
It follows that for each state $|\Psi_{\{q_\alpha\} }\rangle$ in the chosen sector we have 
\begin{equation}
-\hat{\tau}_{j-1,\,j}^x\hat{\tau}_{j,\,j+1}^x \hat{\xi}_j^z |\Psi_{\{q_\alpha\} }\rangle = q_j|\Psi_{\{q_\alpha\} }\rangle \ \ \ \forall j.
\end{equation}
%{The state $|\Psi\rangle$ can thus be expressed as
%\begin{equation}
% |\Psi\rangle=\sum_{\{q_\alpha\}}\left(\prod_\alpha P_\alpha(\{q_\alpha\})\right)|%\Psi\rangle=\sum_{\{q_\alpha\}}p_{\{q_\alpha\}}\Psi_{\{q_\alpha\}}
%\end{equation}
%}
In the second step, we exploit the above relation in order to cancel the matter field operators $\hat{\xi}_j^\alpha$ from  the Hamiltonian $\hat H(t)$.
The derivation is now straightforward. First we have
\begin{align}
 &\, \frac{m}{2} \hat{\xi}_j^z+J\sum_{j=2}^{L-1}\hat{\tau}_{j-1,\,j}^x\hat{\tau}_{j,\,j+1}^x\\ \nonumber
 =&\, -q_j \frac{m}{2} \hat{\tau}_{j-1,\,j}^x\hat{\tau}_{j,\,j+1}^x+J\hat{\tau}_{j-1,\,j}^x\hat{\tau}_{j,\,j+1}^x\\ \nonumber
 =&\, J(1-q_j \frac{m}{2J}) \hat{\tau}_{j-1,\,j}^x\hat{\tau}_{j,\,j+1}^x \ \ \ \forall j\,.
 \end{align}
 Then, we redefine the operators $\hat{\tau}_{j,\,j+1}^\alpha$ in order to cancel the matter field from the second part of $\hat H(t)$
 \begin{align} \nonumber
{\tau^ \prime}_{j,\,j+1}^x &= \tau_{j,\,j+1}^x\\
{\tau^ \prime}_{j,\,j+1}^y &= \xi_j^x\tau_{j,\,j+1}^y\xi_{j+1}^x\equiv\tau_{j,\,j+1}^y \\  \nonumber
{\tau^ \prime}_{j,\,j+1}^z &= \xi_j^x\tau_{j,\,j+1}^z\xi_{j+1}^x\equiv\tau_{j,\,j+1}^z\,.
\end{align}
Note that the proper commutation relations are still satisfied, in particular $\left({\tau^ \prime}_{j,\,j+1}^z\right)^2 =\mathbf{1}$ and $
[{\tau^ \prime}_{j,\,j+1}^z,{\tau^ \prime}_{k,\,k+1}^z] =0 \ \ \  \forall\,k,\,j$. By applying this substitution to the Hamiltonian $\hat H(t)$ we have that
\begin{equation}
h(t)\hat{\xi}_j^x\hat{\tau}_{j,\,j+1}^z\hat{\xi}_{j+1}^x=h(t){\tau^ \prime}_{j,\,j+1}^z  \ \ \ \forall j\,, 
\end{equation}
where the prime will be henceforth omitted. The same substitution allows to cancel out the matter field in the term $\hat H_K$, which is introduced in Eq.~(4) of the main paper as a function of the gauge field only.
{We thus obtain the effective Hamiltonian (Eq.~(3) of the main text) for the superselection sector defined by the static charges $\{q_\alpha\}$. Gauge invariant observables can now be computed by summing over the sectors
\begin{multline}
 \langle \Psi|\hat O(t)|\Psi\rangle=\sum_{\{q_\alpha\}}p_{\{q_\alpha\}}\langle\Psi_{\{q_\alpha\}}|\hat O(t)|\Psi_{\{q_\alpha\}} \rangle\\
 \sum_{\{q_\alpha\}}p_{\{q_\alpha\}}\langle\psi(t)_{\{q_\alpha\}}|\hat O'_{\{q_\alpha\}}|\psi(t)_{\{q_\alpha\}}\rangle
\end{multline}
where $p_{\{q_\alpha\}}=||P_{\{q_\alpha\}}|\Psi\rangle||^2$ gives the projective probability of the initial state $|\Psi\rangle$ on the sector $\{q_\alpha\}$,  $\hat O'_{\{q_\alpha\}}$ is the operator obtained from $\hat O$ after integration of the matter field, and $|\psi(t)_{\{q_\alpha\}}\rangle$ is the state of the gauge field evolved with the Hamiltonian $\hat H_{\{q_\alpha\}}(t)$. As stated in the main text, we can now treat the originally translation-invariant model by computing averages of an effective model with a quenched disorder characterized by a probability distribution $p_{\{q_\alpha\}}$. For simplicity (although not necessary), we choose a class of initial states with the property that $p_{\{q_\alpha\}}=1/\mathcal{N}$, i.e. with uniform weights over all the sectors. We now prove that with the choice of initial states reported in the main text this property is indeed satisfied.}

{
The state for the Higgs field is a product state of spins, each one living on the equator of the Bloch sphere ($\langle\xi^z_j\rangle=0$ for every $j$). For each $j$ we can find
the unit vector $\hat n_j=\cos\theta_j \hat x+\sin\theta_j \hat y$ giving its position on the Bloch sphere, such that
\begin{equation}
 \vec \xi_j\cdot \hat n_j|\varphi\rangle_H=|\varphi\rangle_H.
\end{equation}
We now consider a generic sector $\{q_\alpha\}$ and we see that for every site $i$ we have
\begin{multline}
 (\vec \xi_i\cdot \hat n_i)\left(\prod_j P_j(q_j)\right)|\Psi\rangle
 =\left(\prod_{j\neq i} P_j(q_j)\right) P_i(-q_i)(\vec \xi_i\cdot \hat n_i)|\Psi\rangle
 \\=\prod_{j\neq i} P_j(q_j) P_i(-q_i)|\Psi\rangle
\end{multline}
where we used the fact that $\{(\vec \xi_i\cdot \hat n_i), \hat \xi_i^z\}=0$. From the unitarity of $(\vec \xi_i\cdot \hat n_i)$ we derive the relation between the norms
\begin{multline}
 \norm{\left(\prod_j P_j(q_j)\right)|\Psi\rangle}=\norm{(\vec \xi_i\cdot \hat n_i)\left(\prod_j P_j(q_j)\right)|\Psi\rangle}\\
 =\norm{\left(\prod_{j\neq i} P_j(q_j)\right) P_i(-q_i)|\Psi\rangle},
\end{multline}
which implies that the projections on sectors which only differ for one local charge have equal norms.
Since the last relation holds for every set of $\{q_\alpha\}$ and for every $i$, we find that all the probabilities $p_{\{q_\alpha\}}$ have to be equal, with $p_{\{q_\alpha\}}=1/\mathcal{N}$.
}

%.................................................................................................................................................................%
\section{Single-particle Floquet spectrum} \label{single:sec}
\begin{figure}[b]
  \begin{center}
    \includegraphics[width=8cm]{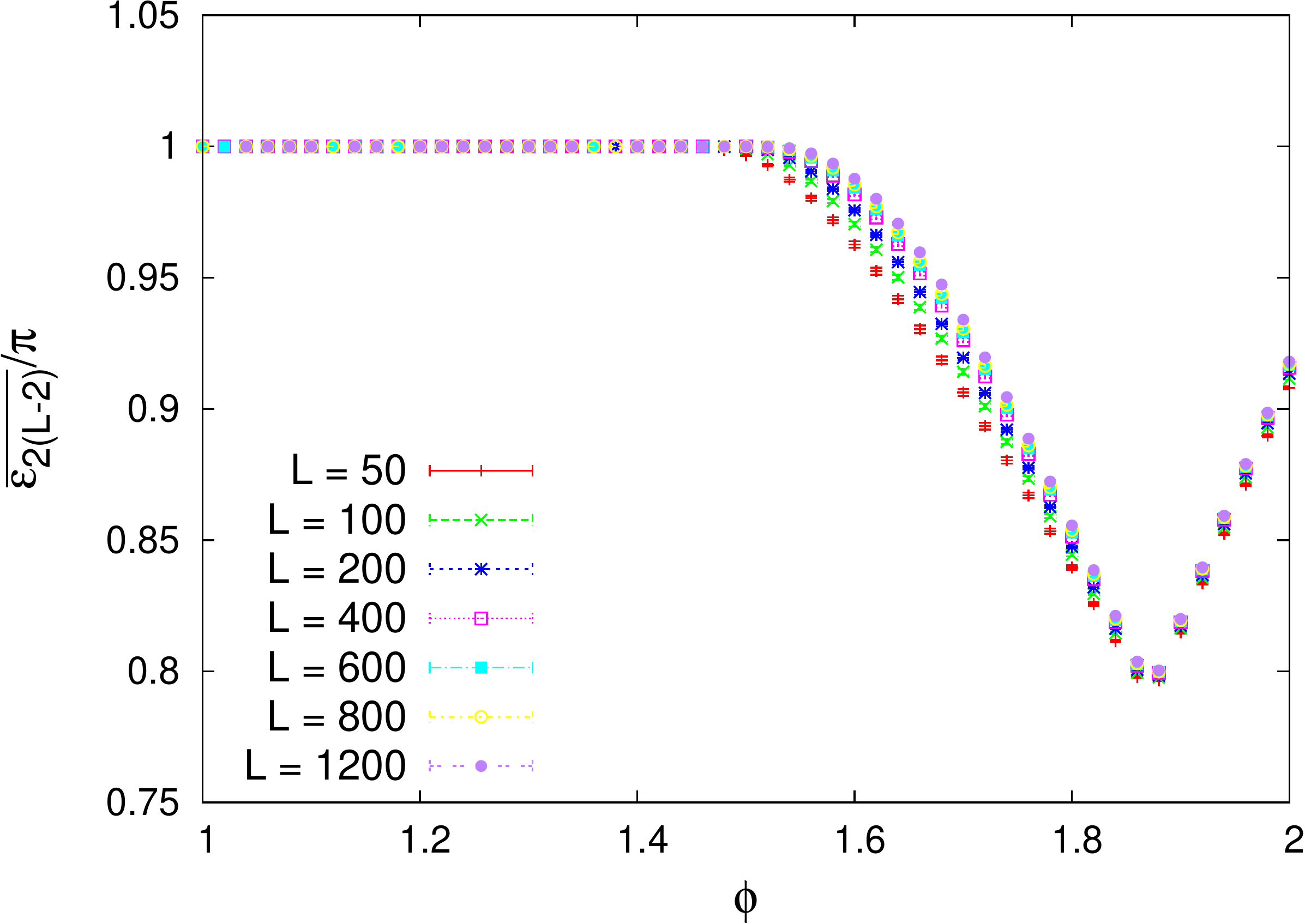}
\caption{ $\epsilon_{2(L-2)}$ averaged over pseudo-disorder versus $\phi$ for different values of the system size $L$. Numerical parameters: $K=0$, $J=1.0$, $h/J=0.2$, $m/J=0.5$, $T=1.0$, $N_{\rm real}\ge104$, open boundary conditions, initial state with $h_0=0.5$ (see main text).}
\label{fig1_app}
  \end{center}
\end{figure}

Fig.~\ref{fig1_app} shows the average over $N_{\rm real}$ pseudo-disorder realizations of $\epsilon_{2(L-2)}$. We can see that it stays at $\pi$ in an interval of $\phi$ larger than the one where $\overline{\delta_\pi}$ is nonvanishing (see Fig.~3 of the main paper). 

\begin{figure}[b]
  \begin{center}
    \includegraphics[width=8cm]{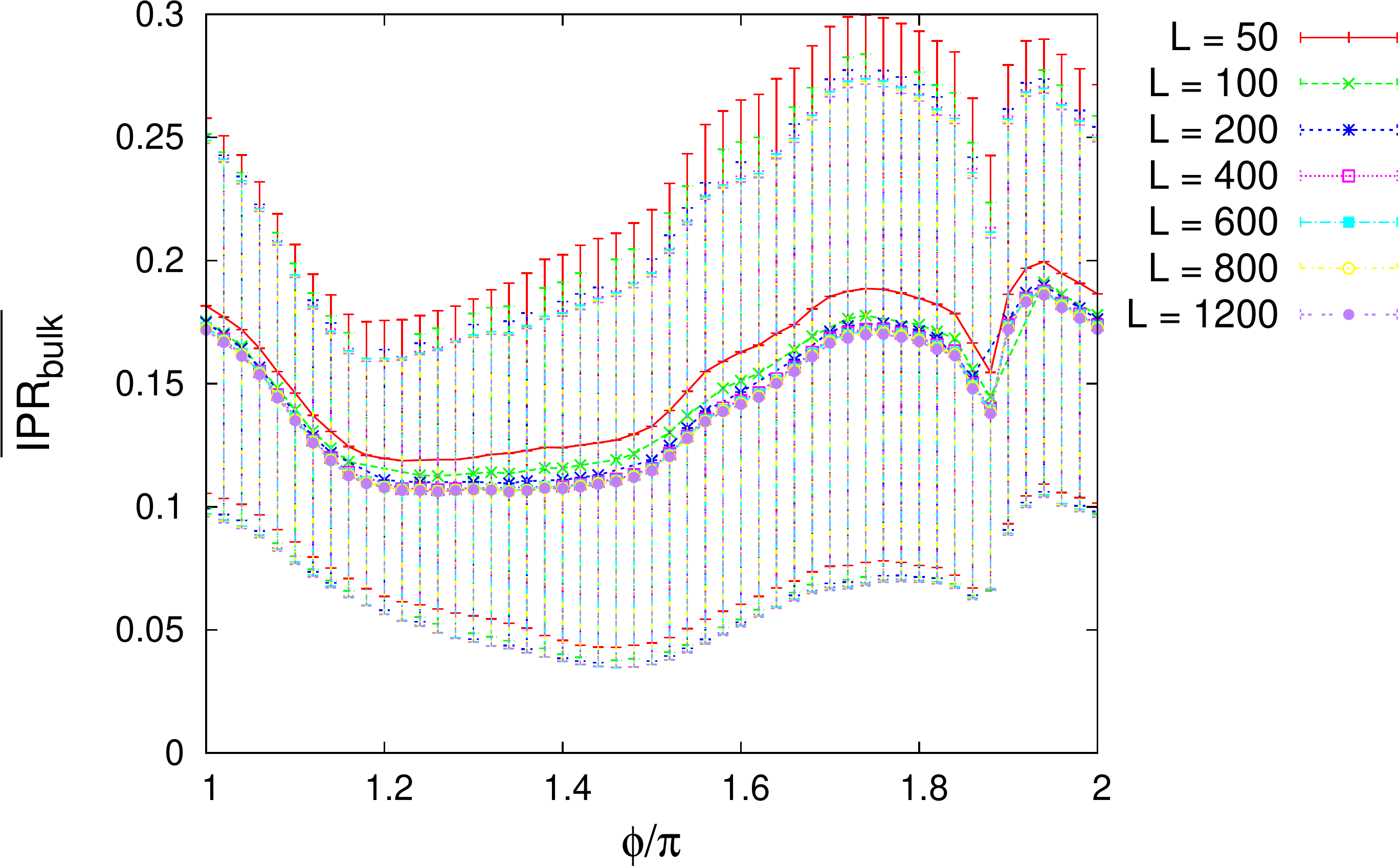}
\caption{ Bulk-averaged single-particle Floquet inverse participation ratio [see Eq.~\eqref{IPRLA:eqn}]. Numerical parameters: $K=0$, $J=1.0$, $h/J=0.2$, $m/J=0.5$, $T=1.0$, $N_{\rm real}\ge104$, open boundary conditions, initial state with $h_0=0.5$ (see main text).}
\label{fig2_app}
  \end{center}
\end{figure}

Fig.~\ref{fig2_app} shows the bulk-averaged single-particle Floquet inverse participation ratio defined as
\begin{align} \label{IPRLA:eqn}
 \overline{\rm IPR_{\rm bulk}}&=\frac{1}{2L-4}\sum_{\alpha\in{\rm bulk}}\overline{{\rm IPR}_{\alpha}}\quad{\rm with}\nonumber\\
 {\rm IPR}_{\alpha} &= \sum_{j=1}^{L-1}\left(|u_{j\,\alpha}(0)|^4+|v_{j\,\alpha}(0)|^4\right)\,,
\end{align}
where we define $\overline{\cdots}$ as the average over the $N_{\rm real}$ realizations of pseudo-disorder, ${\bf w}_\alpha=\left(\begin{array}{cccccccccc}u_{1,\,\alpha}&\cdots&u_{L,\,\alpha}&|&v_{1,\,\alpha}&\cdots&v_{L,\,\alpha}\end{array}\right)$ are the single-particle Floquet states (see for instance~\cite{emanuele} for more details) and $\alpha$ runs over the $2(L-2)$ values corresponding to the bulk single-particle Floquet states (the ones with $\epsilon_\alpha\neq\pm\pi$). For all the considered values of $\phi$ we can clearly see that $\mean{\rm IPR}_{\rm bulk}$ does not scale with the system size and the same occurs for the error bars (evaluated as the r.~m.~s. fluctuation over the pseudo-disorder realizations). This marks the fact that all the single-particle Floquet states are localized and therefore the model shows Anderson localization for all the considered values of $\phi$.
%.....................................................................................................................................................%
\section{Convergence to the Floquet diagonal ensemble} \label{convergence:sec}
\begin{figure}[b]
  \begin{center}
    \includegraphics[width=8cm]{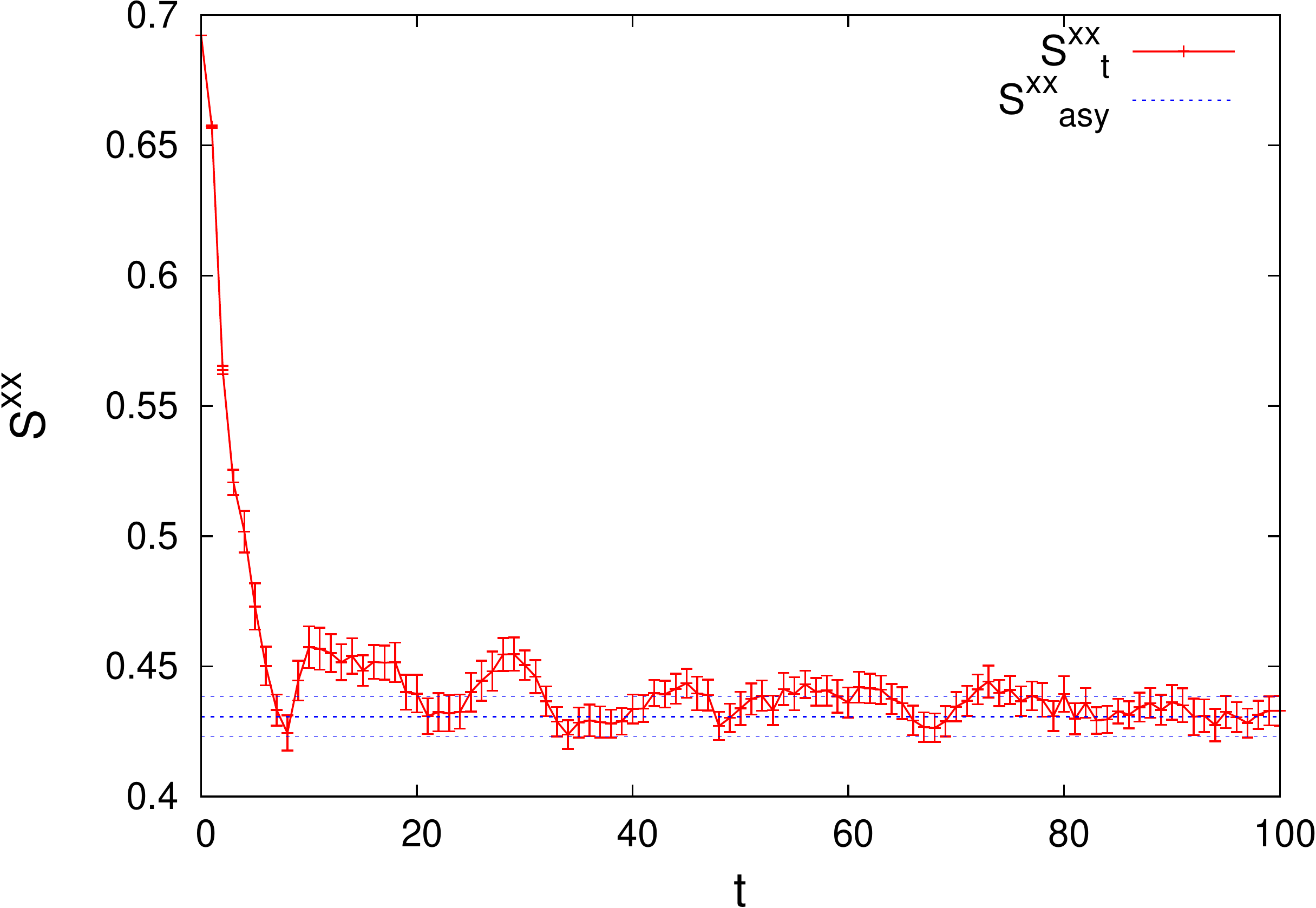}
\caption{ Example of convergence in time of $\mathcal{S}^{xx}_t$ to the Floquet diagonal ensemble value $\mathcal{S}^{xx}_{\rm asy}$. Numerical parameters: $J=1.0, \phi=1.02\pi$, $h/J=0.2$, $m/J=0.5$, $T=1.0$, $N_{\rm real}=48$, open boundary conditions, initial state with $f=0.8$.}
\label{convergence:fig}
  \end{center}
\end{figure}

In the text we have claimed that the correlation order parameter converges for long times towards the Floquet diagonal ensemble value given by Eq.~6 of the main text. We have numerically verified this point; we show an example of this convergence in Fig.~\ref{convergence:fig}. 
We would like also to better discuss this point from the theoretical point of view. Let us expand $\mathcal{S}^{xx}_t$ in the Floquet basis, we find
{\small
\begin{equation}\label{asytime:eqn}
  \mathcal{S}_{t}^{xx} = \sum_{i,j=1,(i\neq j)}^{L-1}\sum_{\footnotesize \begin{array}{c}\beta,\gamma\\\sigma,\sigma'\end{array}}
   \frac{\overline{{R_\beta^\sigma}^*R_\gamma^{\sigma'}{_g\!}\bra{\phi_\beta^\sigma}\tau_{j,\,j+1}^x\tau_{i,\,i+1}^x\ket{\phi_\gamma^{\sigma'}}_g\nep^{-i(\mu_\gamma^\sigma-\mu_\beta^{\sigma'})t}}}{(L-1)(L-2)}\,.
\end{equation}
}
The indexes $\sigma$ and $\sigma'$ can take values $+$ and $-$ and mark the $\mathbb{Z}_2$ symmetry sector. The system is symmetric under the $\mathbb{Z}_2$ symmetry, so the Floquet states are doubly degenerate $\mu_\beta^+=\mu_\beta^-$ $\forall\beta$. Moreover, also the operators $\tau_{j,\,j+1}^x\tau_{i,\,i+1}^x$ are symmetric under this symmetry and we have therefore $_g\bra{\phi_\beta^+}\tau_{j,\,j+1}^x\tau_{i,\,i+1}^x\ket{\phi_\beta^+}_g={_g}\bra{\phi_\beta^-}\tau_{j,\,j+1}^x\tau_{i,\,i+1}^x\ket{\phi_\beta^-}_g$. We can therefore rewrite Eq.~\eqref{asytime:eqn} as
\begin{widetext}
\begin{equation}\label{asytime1:eqn}
  \mathcal{S}_{t}^{xx} = \underbrace{\sum_{i,j=1,(i\neq j)}^{L-1}\sum_{\beta,\sigma}
   \frac{
           \overline{
                |R_\beta^\sigma|^2{_g\!}\bra{\phi_\beta^\sigma}\tau_{j,\,j+1}^x\tau_{i,\,i+1}^x\ket{\phi_\beta^\sigma}_g }
           }
        {(L-1)(L-2)}}_{\textrm{block-diagonal term}}
+
\underbrace{\sum_{i,j=1,(i\neq j)}^{L-1}\sum_{\footnotesize \begin{array}{c}\beta,\gamma\neq\beta\\\sigma,\sigma'\neq\sigma\end{array}}
   \frac{\overline{{R_\beta^\sigma}^*R_\gamma^{\sigma'}{_g\!}\bra{\phi_\beta^\sigma}\tau_{j,\,j+1}^x\tau_{i,\,i+1}^x\ket{\phi_\gamma^{\sigma'}}_g\nep^{-i(\mu_\gamma^\sigma-\mu_\beta^{\sigma'})t}}}{(L-1)(L-2)}}_{\textrm{off-diagonal term}}\,.
\end{equation}
\end{widetext}
The off-diagonal term vanishes in the long time after the disorder average, due to the destructive interference between the oscillating phase factors. Only the block-diagonal term survives. Thanks to the degeneracy of the expectations with respect to the index $\sigma$, we can compute this term directly using the Floquet states given by the numerical diagonalization (which, for each $\beta$, are in general superpositions of $\sigma=+$ and $\sigma=-$). Moreover, $|R_\beta^+|^2+|R_\beta^-|^2$ takes the same value whichever basis in the degenerate Floquet subspace is considered. That's why in the main text we do not write the index $\sigma$.
%
%.....................................................................................................................................................%
\section{No time crystal for continuous gauge symmetry} \label{gauge:sec}
Here we briefly discuss the case of a continuous gauge symmetry. In order to show that the time-crystal question in this case is very delicate (and most probably a discrete time-crystal behaviour in this form is impossible) let
us consider the lattice Schwinger model in the Wilson formulation as an example. As discussed in Ref.~\cite{heyl}, this model displays an extremely slow dynamics, which is qualitatively less ergodic than conventional MBL - in particular, with entanglement entropy growing as $\ln(\ln (t))$. However, due to the absence of any periodicity in the gauge field Hilbert space, it is not possible to identify a clear time-dependent Hamiltonian whose dynamics could lead to a time crystal: for instance, applying the same recipe as above would lead to an infinite period. This case immediately illustrates that the absence of ergodic dynamics -- typical of gauge theories due to superselection sectors -- is by far not enough for engineering translation-invariant time crystals: Identifying the proper gauge symmetry is absolutely key and, in the present context, doable thanks to rather direct analogies with inhomogeneous clock models emerging from a specific class of initial states.

\end{document}